\documentclass[12pt]{article}
\usepackage{graphics,cite}
\begin{document}

\begin{titlepage}

\begin{flushright}
UB-ECM-PF 09/20\\
ICCUB-09-222\\
May 2010
\end{flushright}

\vspace{1.3cm}

\begin{center}
{\Large \bf\boldmath
A Phenomenological Study\\
\vspace{2.5mm}
of Bottom Quark Fragmentation\\ 
\vspace{4.mm}
in Top Quark Decay}
\end{center}

\vspace{5mm}

\begin{center}
{\large G.~Corcella$^{1,2,3}$ and F.~Mescia$^4$}\\

\vspace{5mm}

{$^1${\sl Museo Storico della Fisica e Centro Studi e Ricerche 
E.~Fermi,\\
Piazza del Viminale 1, I-00184 Roma, Italy}}
\vspace{2.5mm}

{$^2${\sl Scuola Normale Superiore,
Piazza dei Cavalieri 7, I-56126, Pisa, Italy}}
\vspace{2.5mm}

{$^3${\sl INFN, Sezione di Pisa, Largo Fibonacci 3,
I-56127, Pisa, Italy}}
\vspace{2.5mm}

{$^4${\sl Universitat de Barcelona,\\
Departamento d'Estructura i Constituents de la Materia (ECM)\\
and Institut de Ciencies del Cosmos (ICC),\\
 Av. Diagonal 647, E-08028, Barcelona, Spain}}

\end{center}

\par \vspace{2mm}
\begin{center}
{\large \bf Abstract}
\end{center}
\begin{quote}
  \pretolerance 10000
Top-quark physics is one of the main fields of investigation 
at the Tevatron accelerator and, ultimately, at the LHC.
We perform a phenomenological analysis of
$t\bar t$ events at hadron colliders, with a focus on
observables relying on bottom-quark
fragmentation in top-quark decay.
In particular, 
we investigate the $B$-lepton invariant-mass distribution in the
dilepton channel and
give an estimate of the contribution of
bottom fragmentation to the Monte Carlo uncertainty 
on the top-quark mass reconstruction.
\end{quote}
\end{titlepage}

\section{Introduction}

Bottom-quark fragmentation in top decay ($t\to bW$)
is one of the main sources of uncertainty in
the measurements of the top-quark properties, such as its mass.
In fact, $b$-quark fragmentation enters in the uncertainty on the
$b$-energy scale, contributing to the Monte Carlo systematics
on the top-mass reconstruction (see, e.g., the top quark analyses from CDF
\cite{cdf,cdf1,lucio} and D0 \cite{d0} at the Tevatron 
accelerator).

Monte Carlo generators, such as the general-purpose HERWIG \cite{herwig} 
and PYTHIA \cite{pythia} codes, implementing hard-scattering processes,
parton showers, hadronization and underlying event,
are widely used to simulate $t\bar t$ events at hadron colliders.
In particular, HERWIG and PYTHIA simulate the
hadronization transition according to the cluster \cite{cluster}
and string \cite{string} models\footnote{As an option, PYTHIA 
allows one to interface its showers to
fragmentation functions, such as the Bowler \cite{bowler} 
or Peterson \cite{peter} models.}, 
respectively, containing a few parameters which 
need to be tuned to the data,
e.g., from LEP or SLD experiments.

In order to estimate the contribution of bottom fragmentation
to the Monte Carlo systematic error on top-quark observables, 
one typically compares the results
yielded by the two codes and varies the
hadronization parameters describing the $b\to B$ transition
within suitable ranges.
At the LHC, it is worthwhile mentioning the study \cite{avto}, where 
it was proposed that one could reconstruct the 
top-quark mass in the dilepton channel, by using the 
decays $B\to J/\psi$ and $J/\psi\to \mu^+\mu^-$, $B$ being 
a $b$-flavoured hadron. The top mass was then fitted from 
the peak value of the $m_{J/\psi\ell}$ or $m_{\mu\ell}$
invariant-mass spectra, $\ell$ being 
a charged lepton in $W$ decay $W\to\ell\nu$. 
This analysis estimates that, in the phase of
luminosity $10^5$~pb$^{-1}$, after setting
suitable cuts on transverse momenta and rapidities of final-state 
leptons, one can reconstruct the top mass with an error 
$\Delta m_t\simeq 1$~GeV.
The contribution of $b$-fragmentation to $\Delta m_t$ is found to be
about 600~MeV and is estimated by using the PYTHIA generator along
with the Peterson fragmentation function\footnote{The Peterson 
fragmentation function reads:
$D(x,\epsilon)=A/[x(1-1/x-\epsilon/(1-x))]$, where $x$ is the hadron
energy fraction, $A$ a normalization constant and $\epsilon$ a parameter
to be tuned to the experimental data.}, varying the $\epsilon$
parameter in the range $(5.0\pm 0.5)\times 10^{-3}$.
At the Tevatron, the recent CDF analysis \cite{lucio}
identifies jets containing a candidate muon from
semileptonic $B$-decays (so-called `soft muon $b$-tagging') and measures 
the top mass by using the invariant mass $m_{\ell\mu}$, with
$\ell$ still coming from $W$-boson decay. The overall Monte Carlo 
uncertainty, due to the modelling of $t\bar t$ production and decay 
in HERWIG and PYTHIA,
including $b$-fragmentation as well\footnote{The uncertainties
due to the treatment of 
initial- and final-state radiation were, however, calculated separately 
from the Monte Carlo systematic error.},
was estimated to be $\Delta m_t\simeq 2.1$~GeV \cite{lucio}.

From the point of view of Monte Carlo generators, however, 
the default parametrizations of both HERWIG and PYTHIA 
are unable to fit LEP and SLD data on $B$-hadron production
at the $Z^0$ pole \cite{drol}. In Ref.~\cite{drol} 
the cluster and string models were tuned to such data: after the fits, 
PYTHIA managed to describe the $B$-energy spectrum very well,
whereas HERWIG was only marginally consistent.

Following \cite{drol}, in this paper we wish to perform
a phenomenological study of $t\bar t$
events at hadron colliders, taking particular care about observables
relying on $b$-quark fragmentation in top decay, and
investigate possible discrepancies between HERWIG and
PYTHIA, which may affect the Monte Carlo systematic error on
the top-mass reconstruction.
In particular, our investigation will be especially useful for
the top mass extractions according to Refs.~\cite{avto,lucio},
as those methods strongly depend on the Monte Carlo simulation of the
$b\to B$ transition in top decay.

In Section 2 we shall briefly review the results of Ref.~\cite{drol} on
fitting cluster and string models to LEP and SLD $B$-production data.
In Section 3 we shall present a few results for observables in
$t\bar t$ events depending on the description of $b$-quark fragmentation.
In Section 4, as an example of application of our analysis, we shall
try to estimate the uncertainty on the extraction of $m_t$ in the
dilepton channel from a fit of the invariant-mass $m_{B\ell}$ distribution,
$\ell$ being a lepton from $W$ decay.
In Section 5 we shall summarize the main results of our study and make
some concluding remarks.

\section{Fitting hadronization models to LEP and SLD 
$B$-production data}
In this section we shall shortly summarize the main findings of 
Ref.~\cite{drol}, where the cluster and string models, which simulate
hadronization in HERWIG and PYTHIA, were fitted to LEP and
SLD data on the $B$-hadron spectrum.
Ref.~\cite{drol} considered data on $b$-flavoured hadron production
from the LEP experiments ALEPH \cite{aleph} and OPAL \cite{opal},
and from SLD \cite{sld}. In particular, the ALEPH sample was made
of $B$-mesons, whereas OPAL and SLD also had a small fraction of $B$-baryons,
such as the $\Lambda_b$. 
The $B$ spectrum in $e^+e^-\to b\bar b$ annihilation at the
$Z^0$ pole was studied in terms of
the quantity
\begin{equation}\label{xb}
x_B=\frac{2p_B\cdot p_Z}{m_Z^2}.\end{equation}
In Eq.~(\ref{xb}), $p_B$ and $p_Z$ are the $B$ and $Z^0$ four-momenta, 
respectively. In $Z^0$ rest frame, $x_B=2E_B/m_Z$,  the normalized 
$B$-energy fraction.
As for PYTHIA, it was chosen the scenario with 
parton showers ordered in virtuality,
with an option to reject non-angular-ordered emissions.
Angular ordering is correctly satisfied by the HERWIG cascades
\cite{marweb,marweb1}.

The default parametrizations of HERWIG and PYTHIA 
\footnote{
Ref.~\cite{drol} used HERWIG 6.506
and PYTHIA 6.220.
The latest FORTRAN versions do not actually present any
new features which may change the conclusions of Ref.~\cite{drol}.}
were unable to
acceptably reproduce such data, yielding $\chi^2/\mathrm{dof}=793.4/61$ and
467.9/61, respectively. In \cite{drol}, the two event generators were
therefore fitted to the $x_B$ spectra: the choice of the authors was
to tune only parameters associated with hadronization and 
leave unchanged the ones related to hard scattering and parton showers.
As pointed out in \cite{drol}, whenever one fits just one measured quantity, 
such as $x_B$, the risk is that one may spoil the comparison with other
observables, e.g., light-flavour fragmentation.
Therefore, the fits performed in \cite{drol} are not an official
tuning of HERWIG and PYTHIA, but just an attempt to understand whether
the description of heavy-flavour fragmentation could be improved.

\begin{figure}
\centerline{\resizebox{0.60\textwidth}{!}{\includegraphics{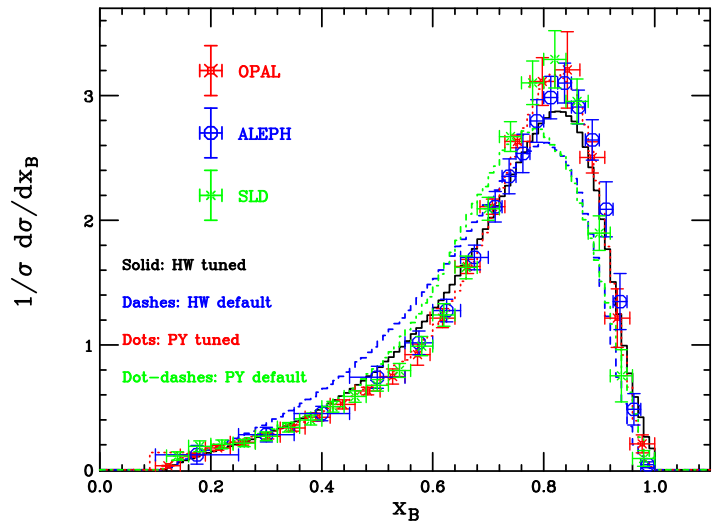}}}
\caption{Data from LEP experiments on the $B$-hadron spectrum in $e^+e^-$
annihilation, 
along with HERWIG and PYTHIA, according to their default versions
and with the hadronization models tuned to such data.}
\label{hpee}
\end{figure}

Table~\ref{tabfit} summarizes the results of
such fits: the $\chi^2$ per degree of freedom refers to all data points,
as if they were coming from one single experiment.
In HERWIG, one fitted CLSMR(1) and CLSMR(2), namely
the Gaussian smearing of the hadron direction with respect to the parent
quark, PSPLT(2), a parameter ruling the mass spectrum of $b$-flavoured
cluster decays, CLPOW, controlling the yield and meson/baryon production,
and DECWT, determining the decuplet/octet ratio.
As for PYTHIA, the three fitted parameters, namely PARJ(41), PARJ(42) and 
PARJ(46), are the $a$, $b$ and $r$ quantities in the Lund/Bowler
fragmentation function \cite{pythia,bowler}:
\begin{equation}
f_B(z)\propto{1\over{z^{1+brm^2_b}}}(1-z)^a\exp(-bm_T^2/z),
\end{equation}
$m_b$ and $m_T$ being the $b$-quark mass and the $B$-hadron transverse mass
respectively.
\begin{table}\small
\begin{center}
\begin{tabular}{|c|c|}\hline
HERWIG  & PYTHIA  \\
\hline\hline
CLSMR(1) = 0.4  (0.0) &                 \\
 \hline
CLSMR(2) = 0.3 (0.0) & PARJ(41) = 0.85 (0.30)\\
\hline
DECWT = 0.7  (1.0)   & PARJ(42) = 1.03 (0.58)\\
\hline
CLPOW = 2.1  (2.0)  & PARJ(46) = 0.85 (1.00)\\
\hline
PSPLT(2) = 0.33 (1.00) &                \\
\hline
\hline
$\chi^2/\mathrm{dof}$ = 222.4/61 (739.4/61) & 
$\chi^2/\mathrm{dof}$ = 45.7/61 (467.9/61)\\
\hline
\end{tabular}
\end{center}
\caption{\label{tabfit}
Best-fit hadronization parameters in
HERWIG and PYTHIA, after comparing with the $B$-hadron energy spectrum
measured at OPAL, ALEPH and SLD,
along with the $\chi^2$ per degree of freedom. In 
brackets, we quote the default values of such parameters.}\end{table}
From Table~\ref{tabfit}, we learn that, after the fit, PYTHIA 
reproduces pretty well the data, while HERWIG is only marginally consistent
with the $x_B$ spectra, although its description of the data is much better
with respect to the default parametrization.
The comparison between data, default and tuned HERWIG and PYTHIA is 
presented in Fig.~\ref{hpee}.
We note that PYTHIA, after the tuning,
gives an excellent description of the data throughout all $x_B$-range, while
the HERWIG prediction, even after fitting the cluster model, is still
below the data around the peak and above the data for middle
values of $x_B$. 
The problems exhibited by HERWIG when its predictions are compared withthe
$x_B$ distributions, have been mostly fixed in the object-oriented version
HERWIG++ \cite{hwpp},
which implements an improved cluster model. In fact, 
Ref.~\cite{gieseke} showed that it is enough tuning the shower cutoff
to obtain a rather good fit of the SLD data. 
Although the employment of HERWIG++ and of the corresponding PYTHIA 8
code \cite{pythiapp}, written in C++,
should be recommended, and not only for the better
description of $B$-hadronization, the FORTRAN versions
of these generators are still widely used. In particular, FORTRAN
HERWIG provides the MC@NLO code \cite{mcnlo}, 
which implements the hard-scattering
process at next-to-leading order (NLO), with parton showers and hadronization.
Moreover, the so-called matrix-element generators, such as, for 
example, the ALPGEN \cite{alpgen} or MadGraph \cite{madgraph} programs, 
often employed to study backgrounds to
$t\bar t$ events, are interfaced to FORTRAN versions of HERWIG and PYTHIA
for showers and hadronization.
It is therefore still useful trying to improve the $b$-fragmentation
sector in the FORTRAN generators and compare their results for a few 
observables relevant to top-quark decay.
The best fits presented in Table~\ref{tabfit} will be the starting
point for the phenomenological analysis which we shall carry out for
$t\bar t$ events at hadron colliders.

Before closing this section, we point out that heavy-quark energy
distributions can also be obtained by using  
resummed calculations, such as Ref.~\cite{cc} for $e^+e^-$ annihilation,
Ref.~\cite{top,topp} for top decays and
Ref.~\cite{gen} for $H\to b\bar b$, $H$ being the Standard Model Higgs 
boson.
Such computations, based on the perturbative-fragmentation formalism
\cite{mele}, resum soft/collinear logarithms with an accuracy which is
usually higher than parton-shower algorithms: therefore, comparisons with
resummations, as done in \cite{drol}, are useful to validate Monte Carlo 
generators and understand the role played by subleading logarithms.
However, resummed computations are too
inclusive to allow a complete investigation of final states.
Also, they still need to be supplemented by phenomenological
non-perturbative fragmentation functions, such as the models
\cite{bowler,peter,kart}, to be comparable with experimental data on
hadron production. In the following, we shall not only 
investigate the $B$-hadron 
energy fraction, but also observables for which resummed calculations
are not currently available. Monte Carlo generators, giving
an exclusive description of final states, are thus the best available
tool to carry out our study.

\section{Top-quark decay observables at hadron colliders}

Given the HERWIG and PYTHIA best fits presented in the above Section,
and relying on the universality of the hadronization transition, we wish
to make predictions for top-decay observables depending on $b$-quark
fragmentation, taking particular care about quantities which 
might be useful for Tevatron or LHC phenomenology.

\subsection{$B$-hadron energy fraction in top decay}
Let us consider top quark decay:
\begin{equation}
t(p_t)\to b(p_b) W(p_W) X(p_X),
\end{equation}
where $X$ stands for extra parton radiation, and the subsequent
transition $b(p_b)\to B(p_B)$. 
A straightforward extension of the $x_B$ variable in Eq.~(\ref{xb}) is
the following quantity:
\begin{equation}\label{xbt}
x_B=\frac{1}{1-m_W^2/m_t^2+m_b^2/m_t^2}
\ \frac{2p_B\cdot p_t}{m_t^2}.\end{equation}
$x_B$ is a Lorentz-invariant variable which corresponds to the
normalized $B$-energy fraction in top rest frame. 
At LEP and SLD, since the $e^+e^-$ collision takes place at the
$Z^0$ pole, the laboratory coincides with the $Z^0$ rest frame.
On the contrary, at the Tevatron or LHC, the laboratory frame is not the
top-quark rest frame, and therefore, in order to measure the $x_B$ quantity,
one would need all four components of top and $B$ momenta.
Such a measurement is obviously not straightforward; however, it is still
interesting 
presenting the $x_B$ spectrum in top decay, in such a way to compare
HERWIG and PYTHIA before and after the fits to $e^+e^-$ data and understand
how much $x_B$ depends on the top quark mass.
We point out that, as already observed in \cite{drol,top}, neglecting
the top width, which is a reasonable approximation as long as 
experimental analyses set cuts of order 10 GeV or higher on the energy of 
final-state jets \cite{orr},
the $x_B$ spectrum is roughly independent of the production
process. Therefore, in such an approximation, our results will
be valid for both Tevatron and LHC: unless stated differently, the 
actual plots
which will shall present are anyway obtained 
running HERWIG and PYTHIA in the LHC environment, i.e. $pp$ collisions
at $\sqrt{s}=14$~TeV.
\begin{figure}[t]
\centerline{\resizebox{0.60\textwidth}{!}{\includegraphics{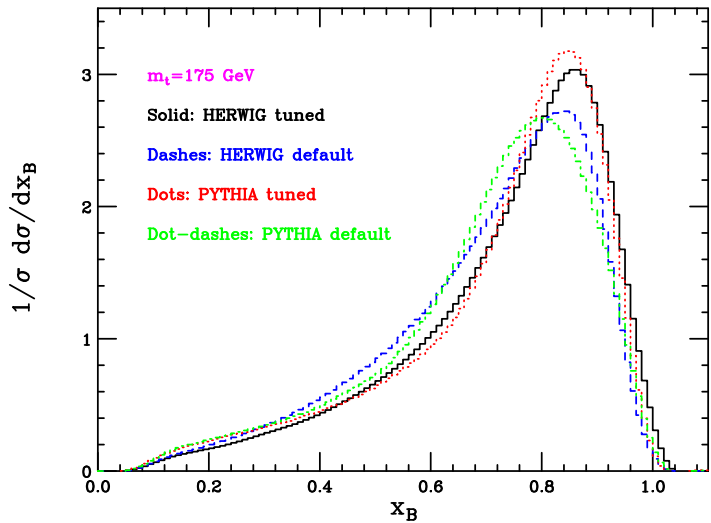}}}
\caption{$B$-hadron spectrum in top
decay, for $m_t=$~175 GeV,
according to HERWIG and PYTHIA, using the default parametrizations
and after fitting cluster and string models to LEP and SLD data.}
\label{hptt}
\end{figure}
\begin{figure}[ht!]
\centerline{\resizebox{0.49\textwidth}{!}{\includegraphics{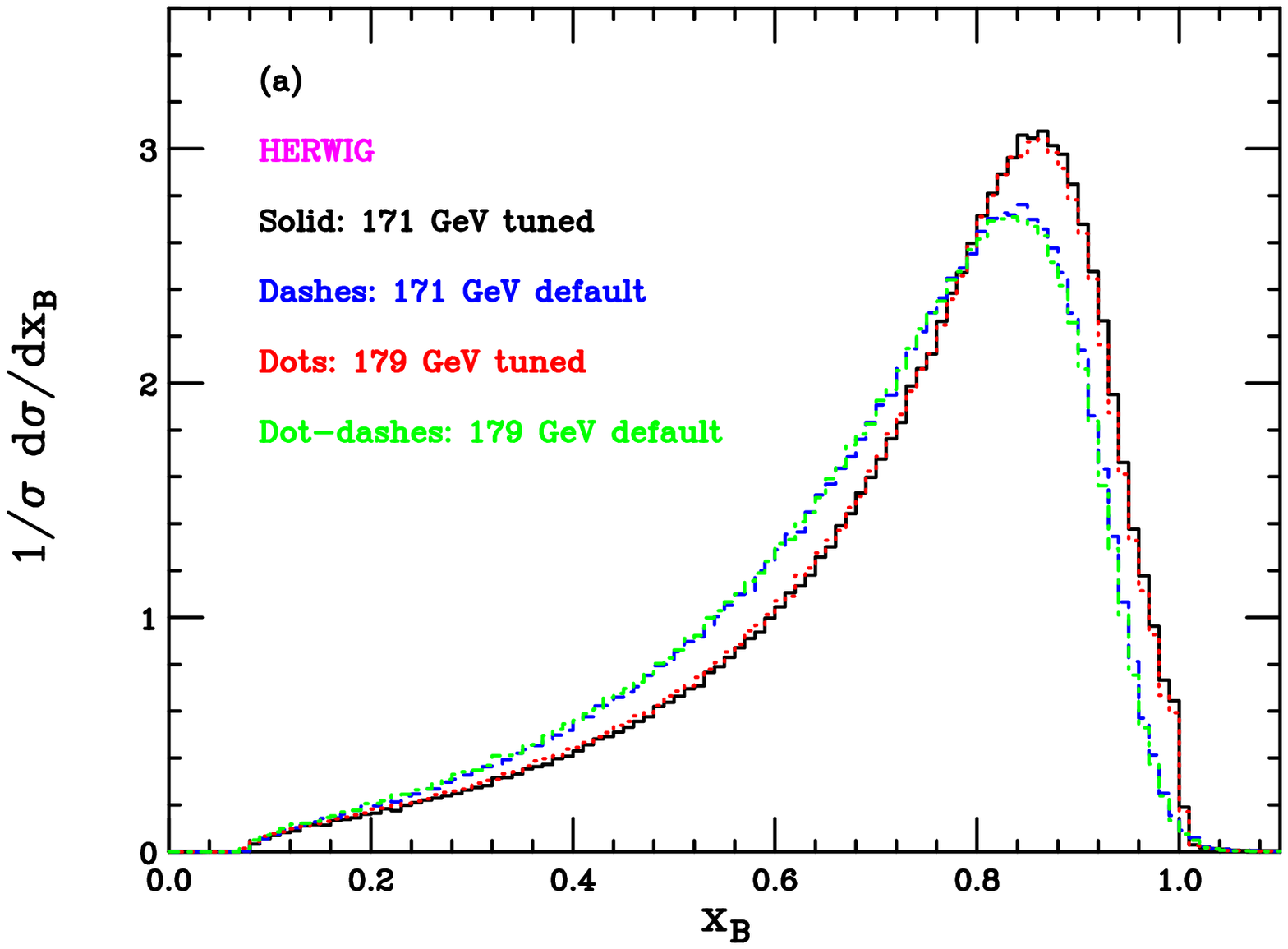}}%
\hfill%
\resizebox{0.49\textwidth}{!}{\includegraphics{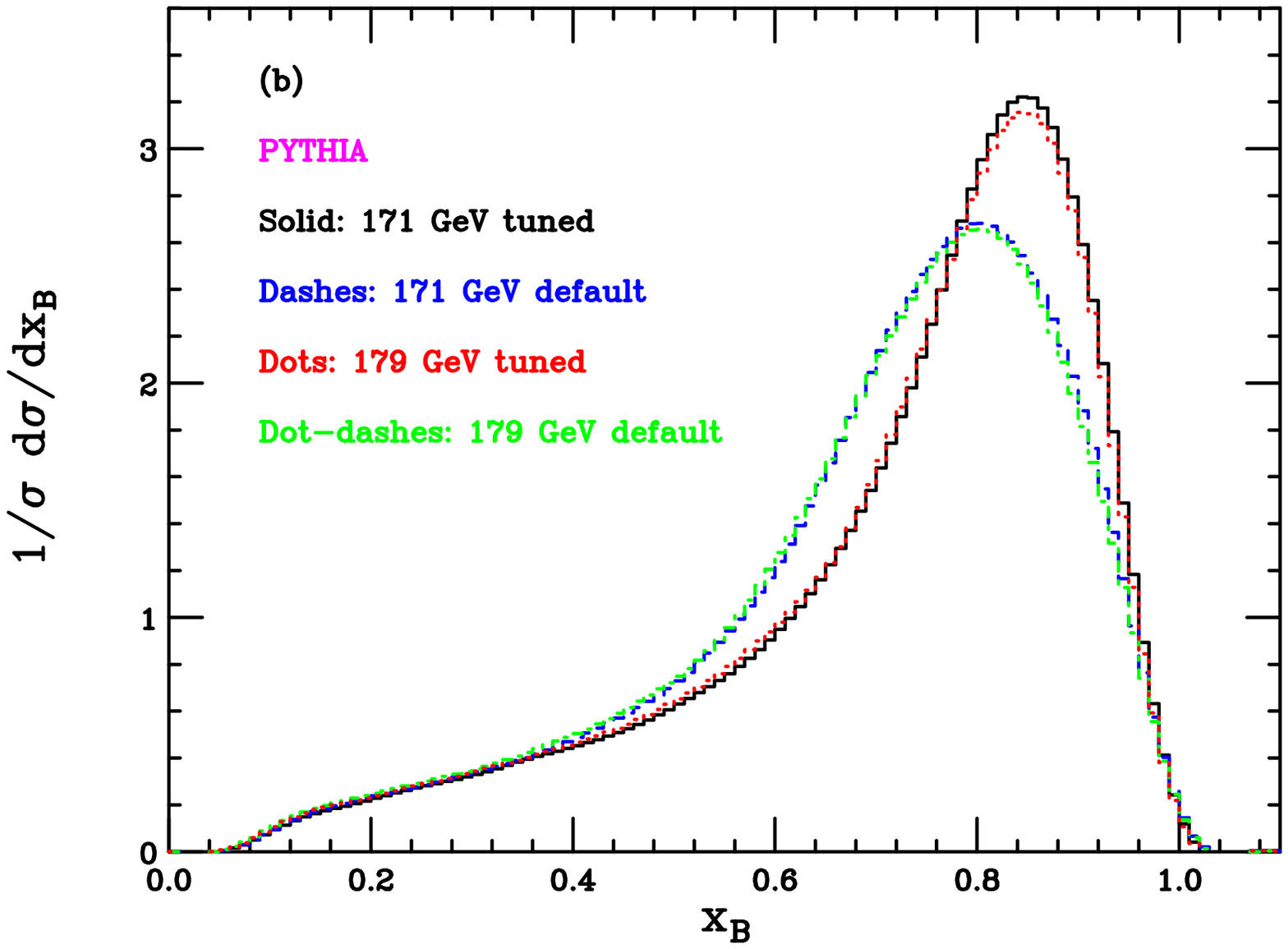}}}
\caption{$x_B$ spectrum in top decay for $m_t=$~171 and 179 GeV,
according to default and tuned HERWIG (a) and PYTHIA (b).}
\label{hpmt}
\end{figure}

Fig.~\ref{hptt} exhibits the $B$-energy distribution in top decay
yielded by PYTHIA and HERWIG, for $m_t=175$~GeV,
using default and tuned parametrizations.
As in the $e^+e^-$ case, the shapes of the spectra are remarkably modified
once we fit string and cluster models: after the tuning, the distributions 
are somewhat narrower and shifted towards higher values of $x_B$.
The comparison of the two
tuned codes is similar to what observed in Fig.~\ref{hpee}: HERWIG yields a
broader distribution and is above PYTHIA for very large and middle values
of $x_B$, whereas it is below PYTHIA around the peak and at small $x_B$.
In Fig.~\ref{hpmt} we present the same spectra, but varying the top mass
from 171 to 179 GeV, and learn that $x_B$ exhibits negligible dependence on 
the top mass, independently of the hadronization
model which one uses.
This is an interesting result: 
if one were able to measure $x_B$, it would be an ideal quantity
to fit $b$-fragmentation parameters, with almost
no dependence on the top-quark mass. 
However, as said above, for the time being, $x_B$ in top decays 
is a difficult
observable to measure.

\subsection{$B$-lepton invariant-mass distribution in the \\dilepton
channel}

In this subsection, we investigate the $B$-lepton invariant-mass ($m_{B\ell}$)
distribution in the dilepton channel, 
where $B$ is a $b$-flavoured hadron coming from top decay
and $\ell$ a charged lepton in $W$ decay
($W\to\ell\nu$). In fact, such a quantity is
closely related to invariant masses 
$m_{\mu\ell}$ and $m_{J/\psi\ell}$,
where the $J/\psi$'s and $\mu$'s 
come from $B$ decays, used in Refs.~\cite{lucio,avto},
to fit the top mass at Tevatron and LHC, respectively.

The $m_{B\ell}$ invariant mass is another boost-invariant 
observable, just relying 
on top decay and not depending on the top-production phase.
Such a quantity was already studied in Refs.~\cite{corcella,cms},
in order to investigate the impact of matrix-element corrections to
simulations of top decays in HERWIG \cite{corsey}. 
Ref.~\cite{cms} also checked that
the $m_{B\ell}$ distribution is roughly the same at the Tevatron and at the 
LHC, 
thus confirming that it is indeed independent of the production mechanism,
which is mainly $q\bar q\to t\bar t$ at the Tevatron and $gg\to t\bar t$
at the LHC. 

\begin{figure}[t]
\centerline{\resizebox{0.60\textwidth}{!}{\includegraphics{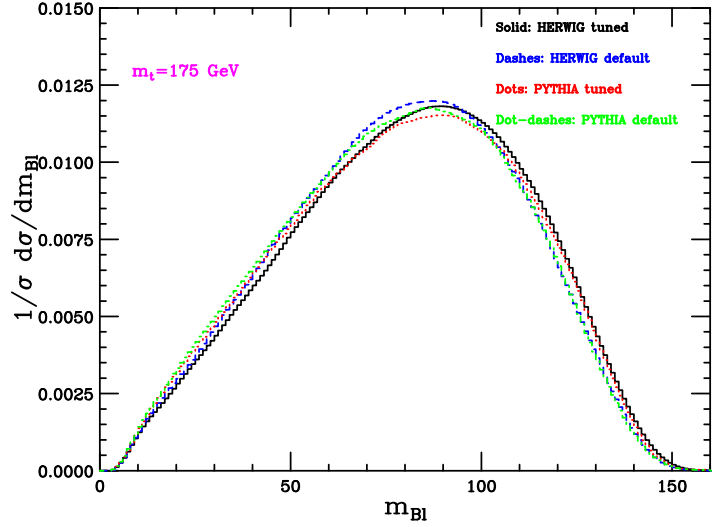}}}
\caption{$B$-lepton invariant-mass distribution, in top decay and in the
dilepton channel, according to tuned and default HERWIG and PYTHIA,
for $m_t=175$~GeV.}
\label{mbl}
\end{figure}
\begin{figure}[ht!]
\centerline{\resizebox{0.49\textwidth}{!}{\includegraphics{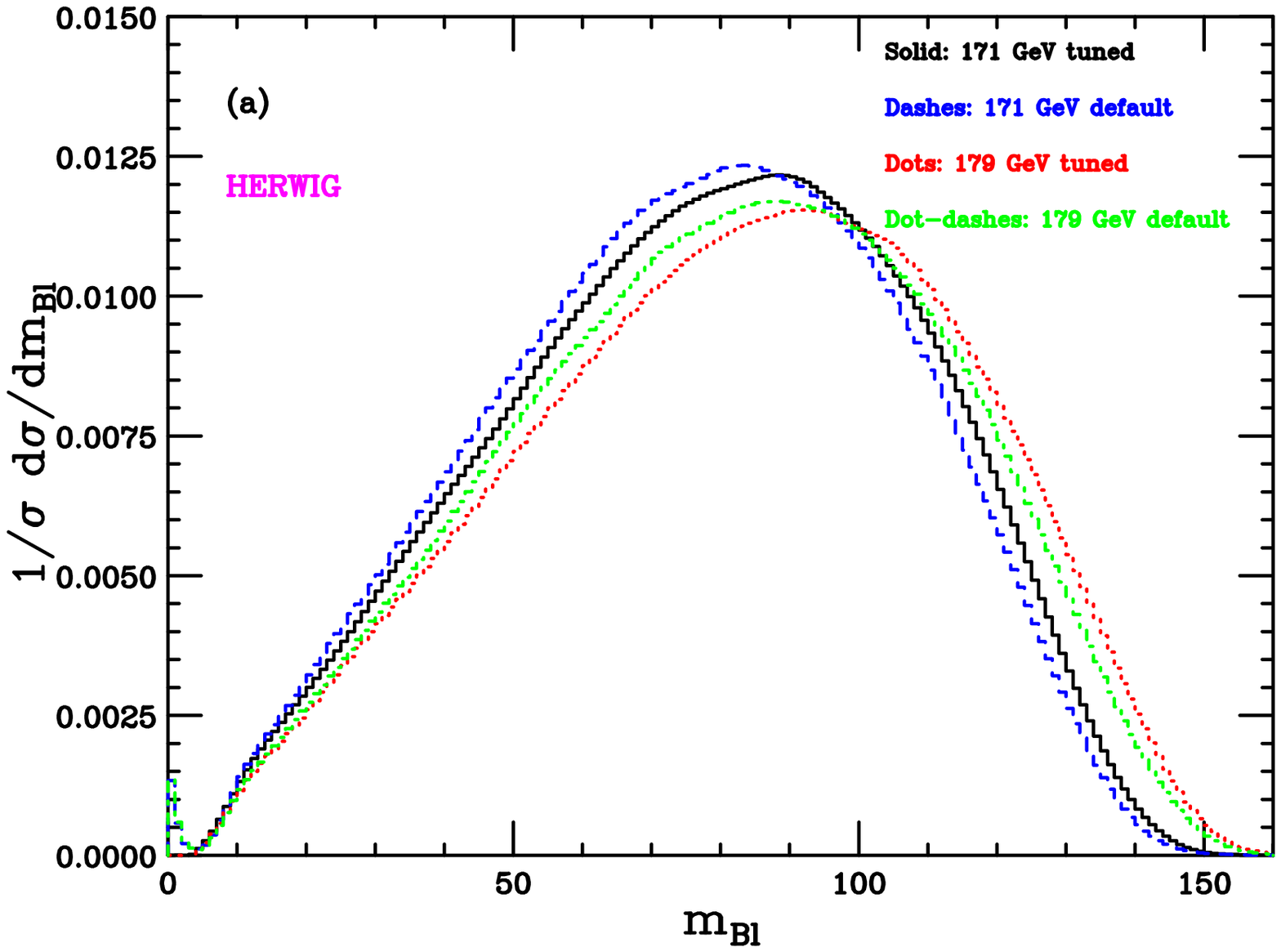}}%
\hfill%
\resizebox{0.49\textwidth}{!}{\includegraphics{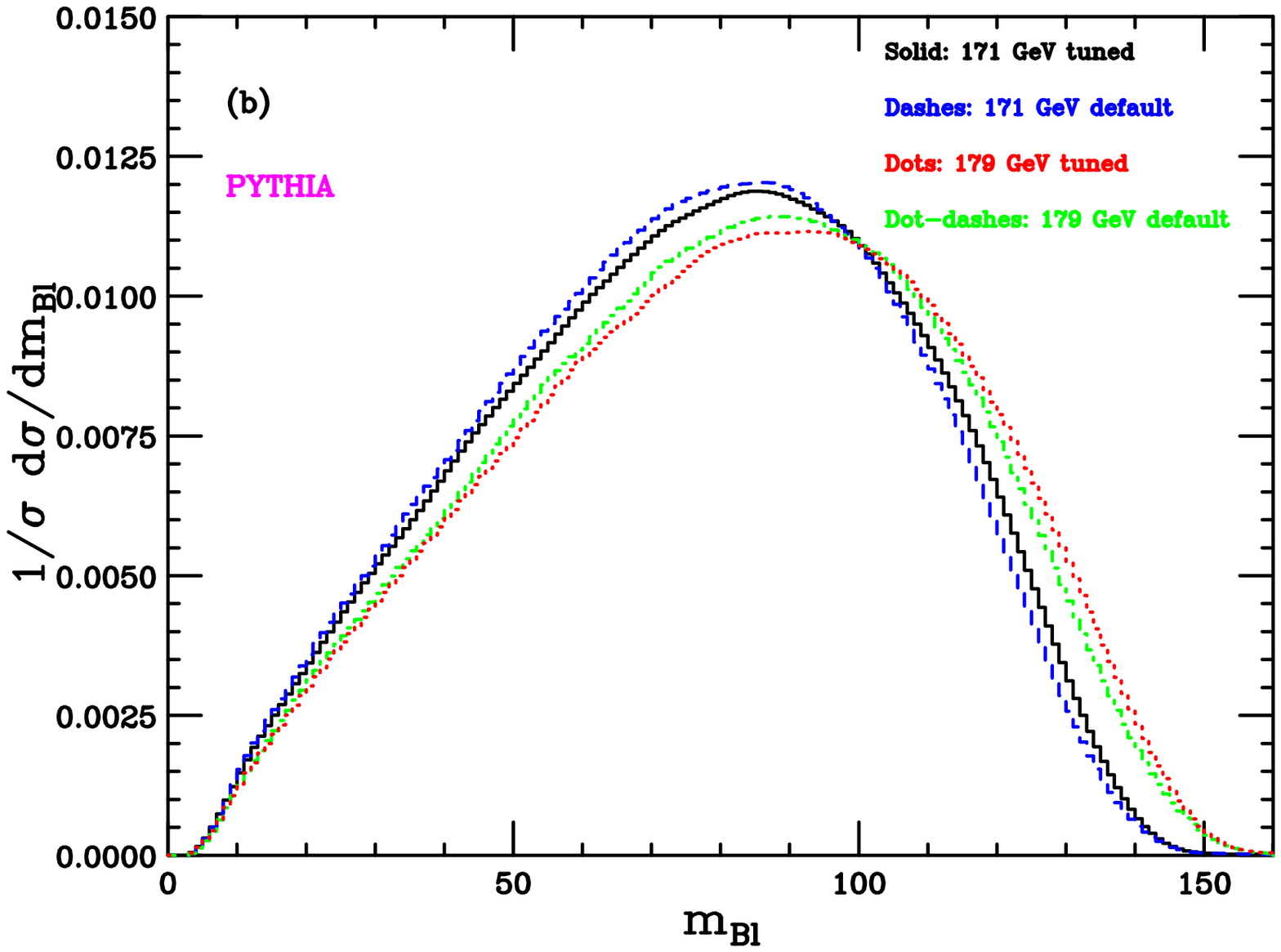}}}
\caption{$m_{B\ell}$ spectrum in top decay for $m_t=$~171 and 179 GeV,
according to default and tuned HERWIG (a) and PYTHIA (b).}
\label{mblmt}
\end{figure}\par
As done in the previous subsection for the $x_B$ quantity, 
we first compare HERWIG and PYTHIA for a given value of
$m_t$ and then we vary the top mass.
Fig.~\ref{mbl} presents the default and tuned $m_{B\ell}$ spectra for 
$m_t=175$~GeV: in both codes, the fit to the $e^+e^-$ data has the effect
to shift the distributions towards larger invariant-mass values
\footnote{The results in Figs.~\ref{mbl} and \ref{mblmt} look
different from the spectra presented in \cite{corcella,cms}, which
were obtained using an unofficial 
preliminary HERWIG version, wherein a few bugs were later found.
The effect of matrix-element corrections to the HERWIG simulation
of top decays found in \cite{corcella,cms} is nevertheless
still confirmed, when using the latest versions, with the bugs fixed.}.
As for the comparison between HERWIG and PYTHIA, the shapes of the
curves yielded by the two generators exhibit visible differences:
HERWIG is above PYTHIA around the peak and below at small $m_{B\ell}$.
At large $m_{B\ell}$ the discrepancy becomes very little, with
HERWIG still giving a slightly higher differential cross section.

Looking at Fig.~\ref{mblmt}, we learn that the behaviours of HERWIG
and PYTHIA spectra with respect to $m_t$ are rather similar, in both
default and tuned versions. Increasing $m_t$ shifts the $m_{B\ell}$ 
spectrum towards higher invariant masses, as one would expect on
physical grounds.
We can anticipate that in Section 4 we shall thoroughly study the
above spectra: we shall compute the Mellin moments and discuss of a possible
extraction of the top mass from a fit of the mean value 
$\langle m_{B\ell}\rangle$.

\subsection{$B$-hadron transverse momentum spectrum}
In this subsection, we investigate the transverse momentum of the
$B$-hadron in top decay $(p_{T,B})$ in the laboratory frame.
Clearly, such an observable is not Lorentz invariant and, unlike
$x_B$ and $m_{B\ell}$, it does not depend only on the decay, but the
production phase is essential in determining its spectrum.
However, it is still useful to study such a quantity: a
measurement of $p_{T,B}$ is more feasible than $x_B$ and, as it happens, e.g., 
for the $W/Z$ transverse momentum in Drell--Yan processes, can be useful
to study the experimental acceptance for top events \cite{sandra}.
In the $p_{T,B}$ spectrum, the hadronization parameters will certainly 
play a role, although, as explained above, a number of other quantities, 
in particular initial-state radiation, are relevant.
\begin{figure}[t]
\centerline{\resizebox{0.60\textwidth}{!}{\includegraphics{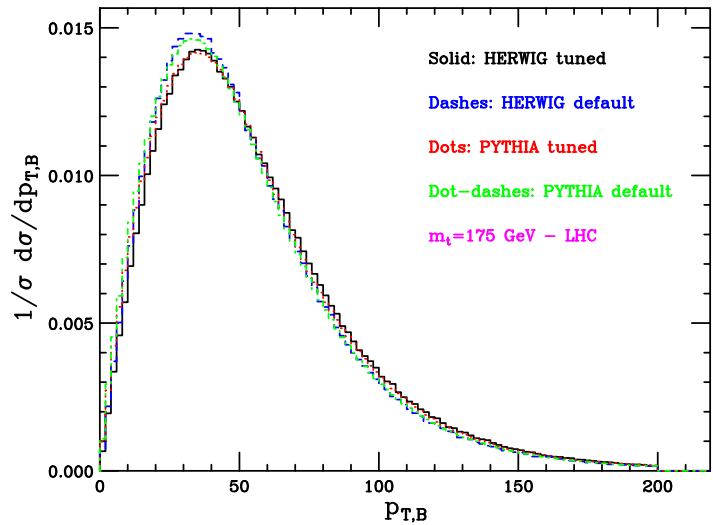}}}
\caption{Transverse momentum distributions of $B$-hadrons in top decay
at the LHC,
according to tuned and default HERWIG and PYTHIA,
for $m_t=175$~GeV. $p_{T,B}$ is evaluated in the laboratory frame.}
\label{pthp}
\end{figure}
\begin{figure}[ht!]
\centerline{\resizebox{0.49\textwidth}{!}{\includegraphics{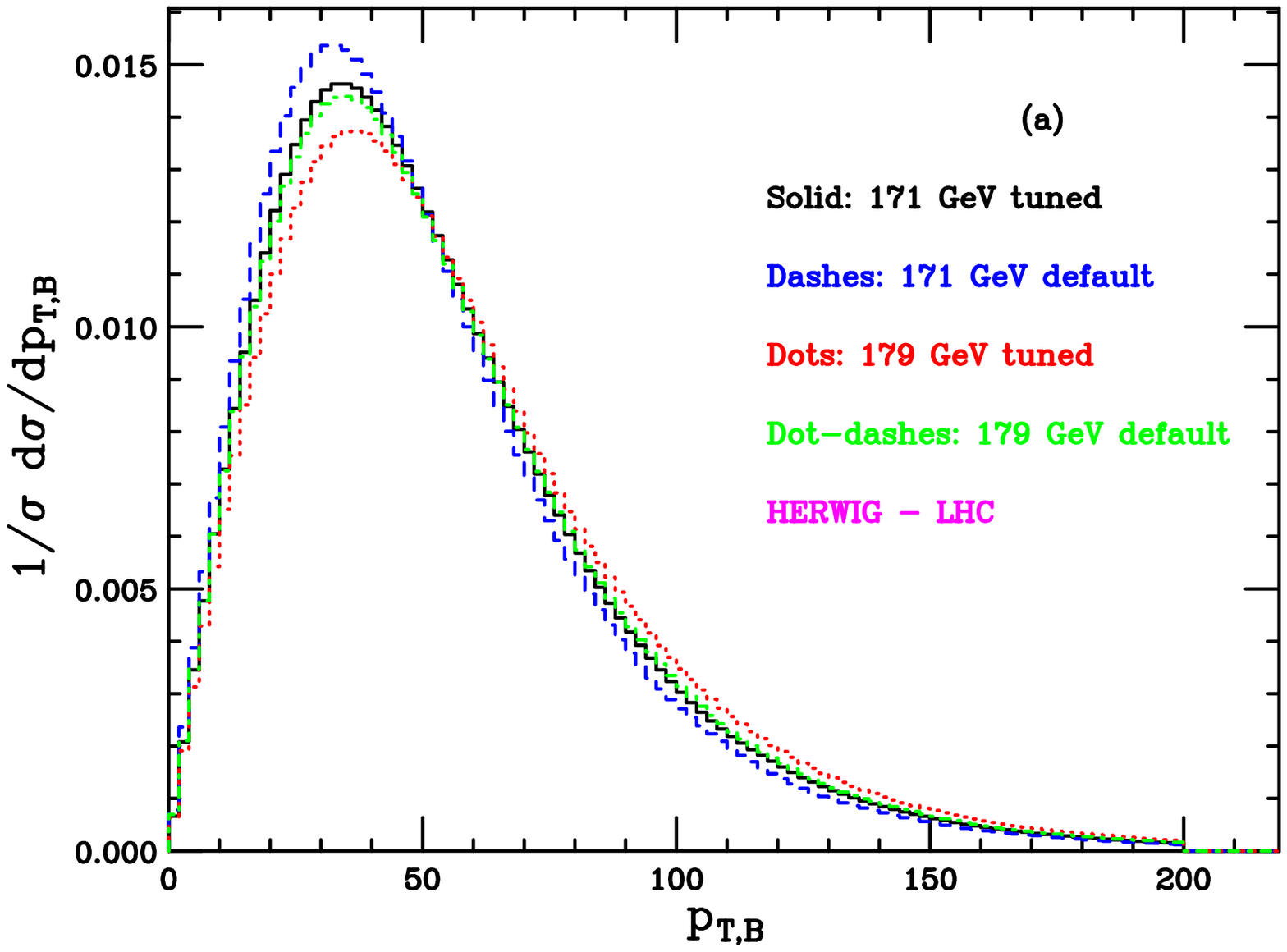}}%
\hfill%
\resizebox{0.49\textwidth}{!}{\includegraphics{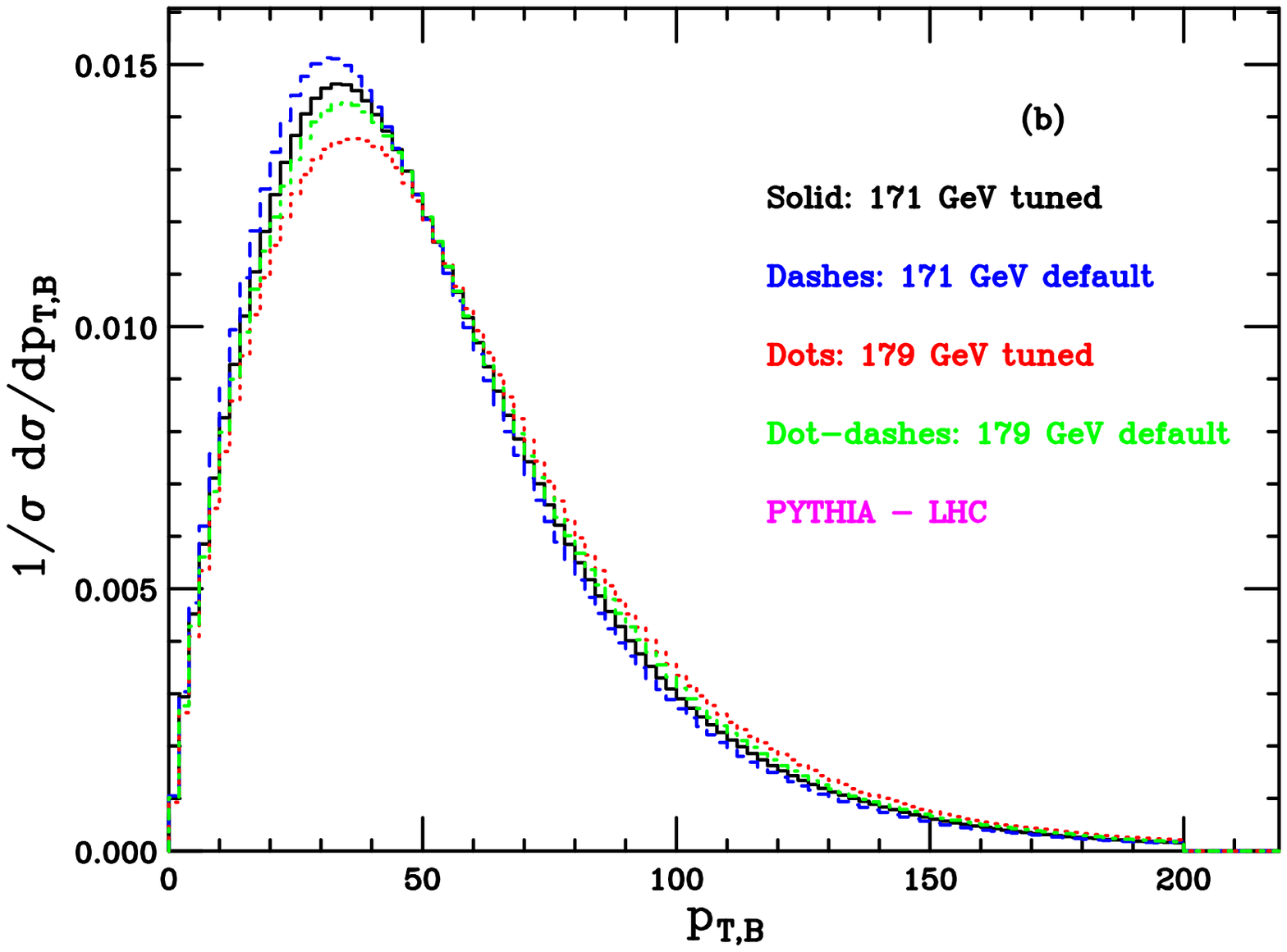}}}
\caption{$p_{T,B}$ spectrum for $m_t=$~171 and 179 GeV,
according to default and tuned HERWIG (a) and PYTHIA (b), at the LHC}
\label{ptmt}
\end{figure} 
\begin{figure}[t]
\centerline{\resizebox{0.60\textwidth}{!}{\includegraphics{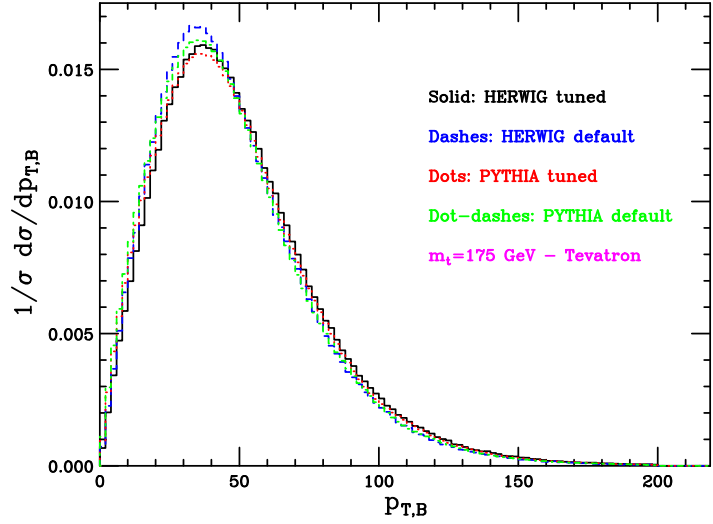}}}
\caption{As in Fig.~\ref{pthp}, but at the Tevatron accelerator.}
\label{ptev}
\end{figure}
\begin{figure}[ht!]
\centerline{\resizebox{0.49\textwidth}{!}{\includegraphics{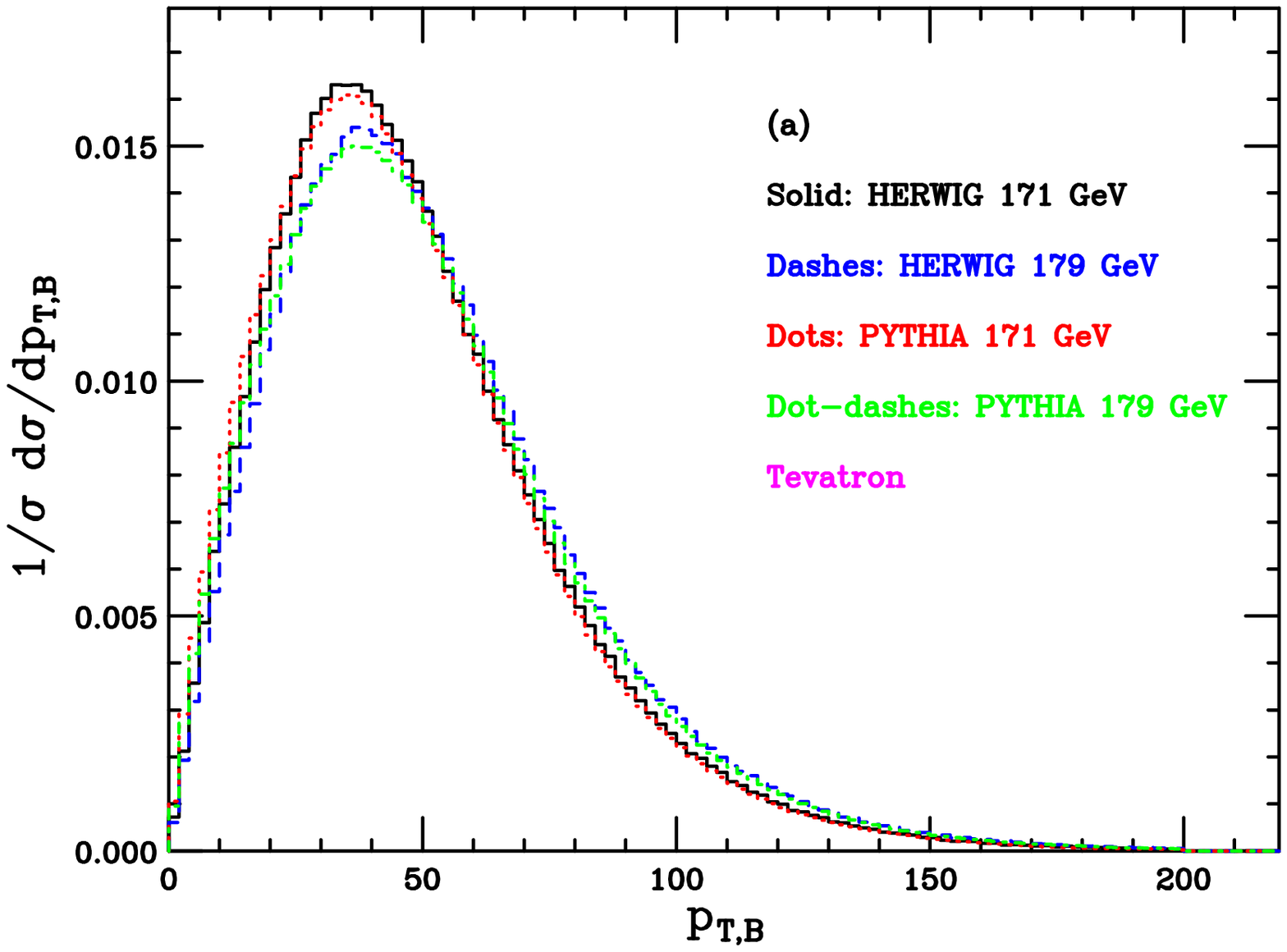}}%
\hfill%
\resizebox{0.49\textwidth}{!}{\includegraphics{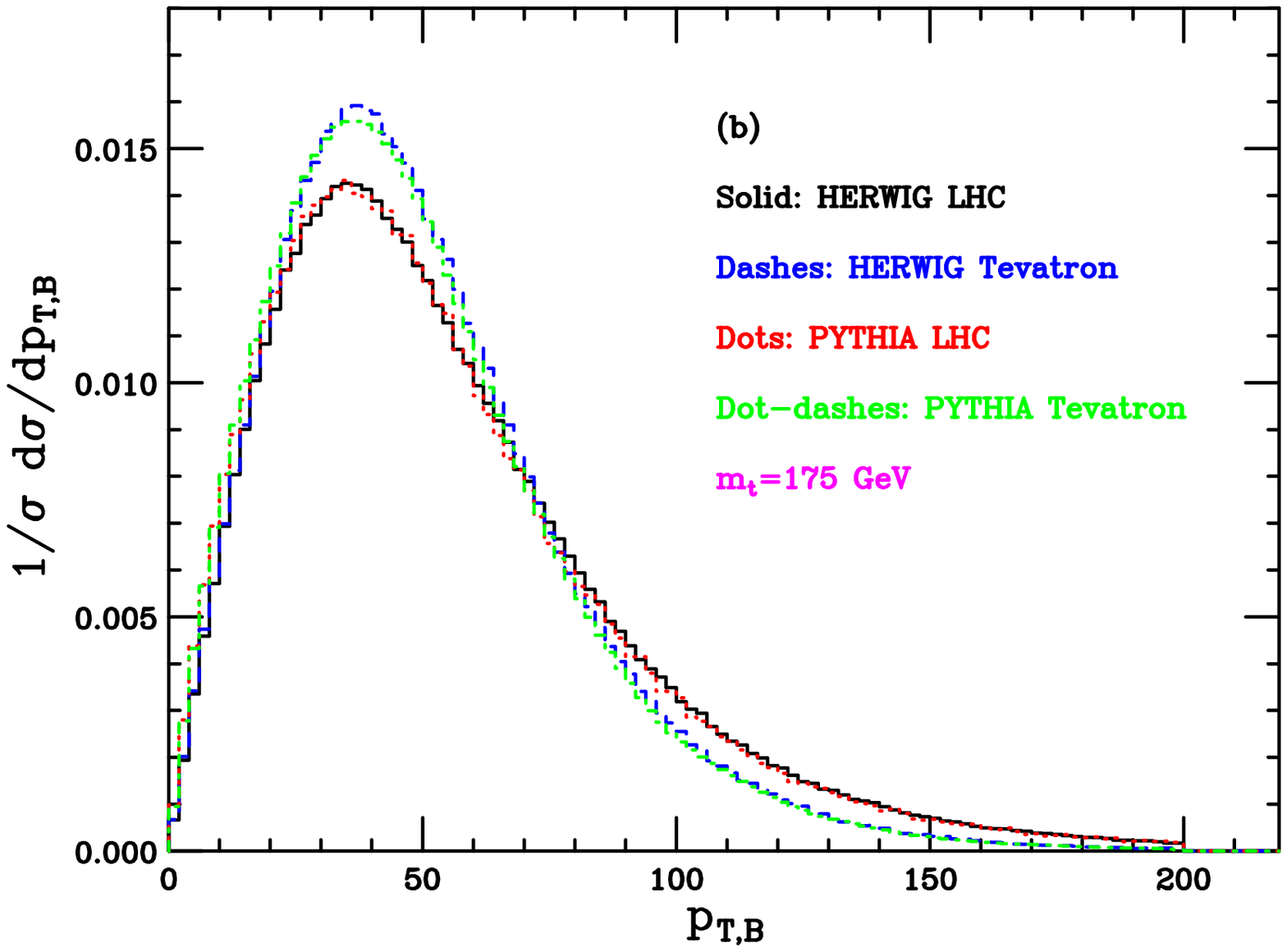}}}
\caption{(a): $p_{T,B}$ spectrum for $m_t=$~171 and 179 GeV,
according to tuned versions of HERWIG and PYTHIA at the Tevatron.
(b): 
Comparison of LHC and Tevatron $B$-hadron transverse momentum spectra,
given by tuned HERWIG and PYTHIA, for $m_t=175$~GeV.}
\label{tevlhc}
\end{figure}\par
In Fig.~\ref{pthp} we present the comparison between HERWIG and
PYTHIA, default and tuned, for $m_t=175$~GeV
at the LHC; in Fig.~\ref{ptmt}
we plot the $p_{T,B}$ distribution for $m_t=171$ and 179 GeV.
As far as this observable is concerned, the discrepancy between
HERWIG and PYTHIA looks smaller than for $x_B$ and 
$m_{B\ell}$, with PYTHIA being slightly above HERWIG at small $p_{T,B}$ and
below for middle-large transverse momenta. 
The effect of the tuning to LEP and SLD data is similar
for both generators: less events at small $x_B$ and at the peak, 
but a higher differential cross section for $p_{T,B}>$~60 GeV.
To be more quantitative, the average value of $p_{T,B}$ for $m_t=175$~GeV
reads $\langle p_{T,B}\rangle\simeq 50.66$ and 50.59 GeV, for default
HERWIG and PYTHIA respectively. After tuning cluster and string
models, HERWIG yields $\langle p_{T,B}\rangle \simeq 53.01$~GeV, 
whereas PYTHIA $\langle p_{T,B}\rangle \simeq 52.20$~GeV.
As for the top-mass dependence of this observable, a higher $m_t$ results
in less events around the peak value, and more with a 
large transverse-momentum $b$-flavoured hadron.

Since we observed that $p_{T,B}$ is not Lorentz-invariant and depends
on the $t\bar t$ production mechanism, it is interesting investigating
such a quantity even at the Tevatron accelerator, i.e. $p\bar p$ 
collisions at 1.96~GeV, where the production mechanism
is mostly $q\bar q\to t\bar t$, whereas gluon-gluon fusion
dominates at the LHC.
In Fig.~\ref{ptev} we present the $B$ transverse-momentum distribution
at the Tevatron, for $m_t=175$~GeV, given by default and tuned HERWIG
and PYTHIA. We learn that at Tevatron energies the effect of the fits 
is qualitatively similar to what found at the LHC: less events are simulated
at small transverse momentum and about the peak, while there are more
top-decay $B$-hadrons at large $p_{T,B}$. 
The mean values are $\langle p_{T,B}\rangle\simeq$
47.91 and 47.66 GeV, according to default HERWIG and PYTHIA,
whereas, when using the parametrization in Table~\ref{tabfit}, 
HERWIG gives $\langle p_{T,B}\rangle\simeq$ 50.23 and PYTHIA
49.02 GeV. Hence, at both Tevatron and LHC,
the two codes are in better agreement
when using the default parametrizations, while the discrepancy gets larger
after the tuning.

In Fig.~\ref{tevlhc} we use instead the two codes with the
fragmentation parameters tuned to the LEP and SLD data:
in Fig.~\ref{tevlhc} (a) we show the $p_{T,B}$
spectrum for $m_t=171$ and 179 GeV, while, for the sake of comparison, 
in Fig.~\ref{tevlhc} (b) we present HERWIG and PYTHIA predictions at the
Tevatron and at the LHC, using $m_t=175$~GeV.
The dependence of the $p_{T,B}$ spectrum on the top mass is like the one
observed at LHC: a higher $m_t$ shifts the $B$ transverse-momentum distribution
towards larger $p_{T,B}$.
As for the comparison Tevatron/LHC, it shows indeed that we are
dealing with an observable depending on the $t\bar t$ production stage and on
the boost from the laboratory frame to the top rest frame, where top
decay is performed. At the Tevatron, due to the lower available energy,
most events are simulated for $p_{T,B}<70$~GeV; 
at the LHC, the large-$p_{T,B}$ tail is instead 
more relevant with respect to the Tevatron. 

\section{Extracting the top mass from the $m_{B\ell}$ spectrum}
As the $B$-lepton invariant mass is a Lorentz-invariant quantity, depending
only on top decay and visibly sensitive to the top mass (see
Fig.~\ref{mblmt}), we can think of using $m_{B\ell}$ 
to fit $m_t$. Also, after convoluting $m_{B\ell}$
with the $B\to J/\psi$ or $B\to\mu$ 
spectra, one will obtain the $m_{J/\psi\ell}$ and $m_{\mu\ell}$
distributions, employed 
in Refs.~\cite{avto,lucio} to extract $m_t$ at the LHC and at the Tevatron,
respectively.

Ideally, if we had data on $m_{B\ell}$, we may directly use them
to validate the Monte Carlo tools and fit the cluster/string models.
For the time being, 
we try to express the $m_{B\ell}$ spectra in terms of the
top mass by computing
the first few Mellin moments in the range 171~GeV~$<m_t<$~179~GeV.
After observing that the fits to LEP and SLD do have a 
strong impact on top-decay observables depending on $b$-fragmentation,
hereafter 
we shall stick to the best-fit parametrizations quoted
in Table~\ref{tabfit}.

\begin{table}[ht!]\small

\begin{center}
\begin{tabular}{|c||c|c|c|c|}\hline
$m_t$ (GeV) & $\langle m_{B\ell}\rangle$ (GeV)& 
$\langle m_{B\ell}^2\rangle$ (GeV$^2$)&
$\langle m_{B\ell}^3\rangle$ (GeV$^3$)& $\langle m_{B\ell}^4\rangle$ 
(GeV$^4$)\\
\hline
171 & 78.39 & $7.01\times 10^3$ & $6.82\times 10^5$ & $7.02\times 10^8$ \\
\hline
173 & 79.52 & $7.22\times 10^3$ & $7.12\times 10^5$ & $7.43\times 10^8$ \\ 
\hline
175 & 80.82 & $7.45\times 10^3$ & $7.46\times 10^5$ & $7.91\times 10^8$ \\
\hline
177 & 82.02 & $7.67\times 10^3$ & $7.79\times 10^5$ & $8.37\times 10^8$ \\ 
\hline
179 & 83.21 & $7.89\times 10^3$ & $8.13\times 10^5$ & $8.86\times 10^8$ \\
\hline\end{tabular}
\end{center}
\caption{\label{tabhw}
First four moments of the $m_{B\ell}$ spectrum in top decay,
yielded by
HERWIG, after tuning the cluster model to ALEPH, OPAL and SLD data,
for 171 GeV~$<m_t<$~179 GeV. } \end{table}
\begin{table}[ht!]
\begin{center}
\begin{tabular}{|c||c|c|c|c|}\hline
$m_t$ (GeV) & $\langle m_{B\ell}\rangle$ (GeV)& 
$\langle m_{B\ell}^2\rangle$ (GeV$^2$)&
$\langle m_{B\ell}^3\rangle$ (GeV$^3$)& $\langle m_{B\ell}^4\rangle$ 
(GeV$^4$)\\
\hline
171 & 77.17 & $6.85\times 10^3$ & $6.62\times 10^5$ & $6.81\times 10^8$ \\
\hline
173 & 78.37 & $7.06\times 10^3$ & $6.94\times 10^5$ & $7.23\times 10^8$ \\ 
\hline
175 & 79.55 & $7.27\times 10^3$ & $7.25\times 10^5$ & $7.67\times 10^8$ \\
\hline
177 & 80.70 & $7.48\times 10^3$ & $7.56\times 10^5$ & $8.12\times 10^8$ \\ 
\hline
179 & 81.93 & $7.71\times 10^3$ & $7.91\times 10^5$ & $8.61\times 10^8$ \\
\hline\end{tabular}
\end{center}
\caption{\label{tabpy}
As in Table~\ref{tabhw}, but using the PYTHIA event generator.} \end{table}

We present the first four moments yielded by HERWIG and PYTHIA
in Tables~\ref{tabhw} and \ref{tabpy}, respectively \footnote{We remind
that, since we plotted $(1/\sigma) (d\sigma/dm_{B\ell})$ everywhere,
our distributions are normalized to unity.}.
From the comparison, we learn
that HERWIG sistematically yields moments which are larger
than PYTHIA, as was already predictable looking at Fig.~\ref{mbl}.
Furthermore, according to both codes, the moments of the
$m_{B\ell}$ spectrum linearly increase with respect to the top mass.

In order to give an estimate of the Monte Carlo uncertainty due to
modelling $b$-quark fragmentation, 
we perform a linear fit of the average
value $\langle m_{B\ell}\rangle$ in terms of $m_t$, by means of the
least-square method \footnote{Of course, given the numbers in
Tables~\ref{tabhw} and \ref{tabpy}, a linear fit will work even for 
the higher Mellin moments of the $m_{B\ell}$ spectrum.}.
The linear fit works very well and the best fits are the following:
\begin{eqnarray}
\langle m_{B\ell}\rangle_{\mathrm{H}} &\simeq & 
-25.31~\mathrm{GeV} +0.61\  m_t\ \  ,\ \  \delta = 0.043~\mathrm{GeV},\\
\langle m_{B\ell}\rangle_{\mathrm{P}} &\simeq & 
-24.11~\mathrm{GeV} +0.59\  m_t\ \  ,\ \  \delta = 0.022~\mathrm{GeV},
\end{eqnarray}
where $\delta$ is the mean square deviation in the fit and the subscripts
H and P refer to HERWIG and PYTHIA, respectively.
\begin{figure}[t]
\centerline{\resizebox{0.60\textwidth}{!}{\includegraphics{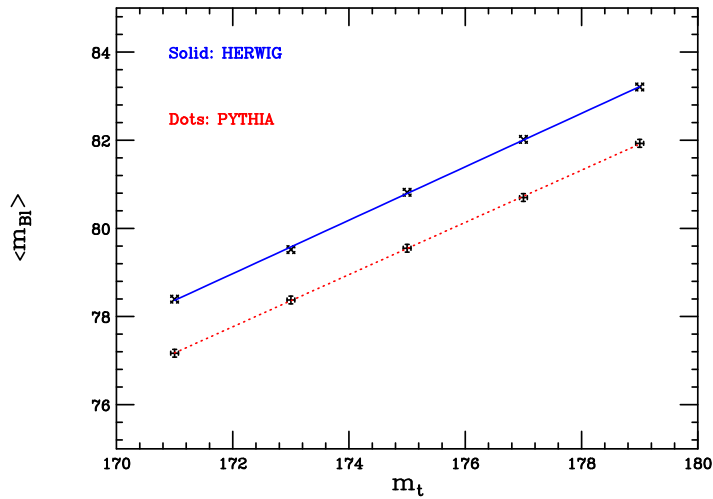}}}
\caption{Linear fits of $\langle m_{B\ell}\rangle$ as a function of 
$m_t$, as obtained from HERWIG and PYTHIA codes. }
\label{fitmt}
\end{figure}
The best-fit straight lines, as a function of $m_t$, are plotted
in Fig.~\ref{fitmt}: we see that, for a given measurement of 
$\langle m_{B\ell}\rangle$, the extracted values of $m_t$ can be quite
different according to whether one uses HERWIG or PYTHIA.
In fact, as Tables~\ref{tabhw} and \ref{tabpy} tell us that 
the typical difference between HERWIG and PYTHIA 
is $\langle m_{B\ell}\rangle\simeq 1.2-1.3$~GeV, the corresponding
uncertainty inferred on $m_t$ can be up to about 
$\Delta m_t\simeq 2$~GeV, given
the slopes of the straight lines in Fig.~\ref{fitmt}.
Such a value of $\Delta m_t$ is clearly quite large, and much above the
600 MeV quoted in \cite{avto}, thus showing that probably such an error,
obtained varying the $\epsilon$ parameter in the Peterson hadronization
model, may have been underestimated. Our $\Delta m_t$ is instead closer
to the 2.1 GeV determined in \cite{lucio} as the Monte Carlo
systematic error when measuring $m_t$ from the $m_{\mu\ell}$ spectrum
at CDF.

Our analysis, however, assumes that one is indeed able to measure the
full $m_{B\ell}$ spectrum, which is obviously quite ideal.
A more realistic estimate of $\Delta m_t$ due to the $b$-quark hadronization
can be obtained if we discard the low- and high-$m_{B\ell}$
tails and restrict ourselves, e.g., to the
range 50~GeV~$< m_{B\ell} <$~120~GeV.
In this range, we obtain the truncated moments of the $m_{B\ell}$ spectrum
presented in Tables \ref{hwt} and \ref{pyt}. The discrepancy between 
HERWIG and PYTHIA is clearly much 
smaller after we cut the tails of the spectrum;
the linear relation of the moments with respect to $m_t$ is nonetheless
still preserved.
\begin{table}[ht!]
\begin{center}
\begin{tabular}{|c||c|c|c|c|}\hline
$m_t$ (GeV) & $\langle m_{B\ell}\rangle$ (GeV)& 
$\langle m_{B\ell}^2\rangle$ (GeV$^2$)&
$\langle m_{B\ell}^3\rangle$ (GeV$^3$)& $\langle m_{B\ell}^4\rangle$ 
(GeV$^4$)\\

\hline
171 & 84.64 & $7.52\times 10^3$ & $6.97\times 10^5$ & $6.70\times 10^8$ \\
\hline
173 & 85.01 & $7.59\times 10^3$ & $7.06\times 10^5$ & $6.81\times 10^8$ \\ 
\hline
175 & 85.43 & $7.66\times 10^3$ & $7.17\times 10^5$ & $6.94\times 10^8$ \\
\hline
177 & 85.78 & $7.72\times 10^3$ & $7.25\times 10^5$ & $7.04\times 10^8$ \\ 
\hline
179 & 86.09 & $7.78\times 10^3$ & $7.32\times 10^5$ & $7.13\times 10^8$ \\
\hline\end{tabular}
\end{center}
\caption{\label{hwt}Truncated moments of the $m_{B\ell}$ spectrum,
according to HERWIG in the range
50 GeV~$<m_{B\ell}<$~120 GeV.} \end{table}
\begin{table}[ht!]
\begin{center}
\begin{tabular}{|c||c|c|c|c|}\hline
$m_t$ (GeV) & $\langle m_{B\ell}\rangle$ (GeV)& 
$\langle m_{B\ell}^2\rangle$ (GeV$^2$)&
$\langle m_{B\ell}^3\rangle$ (GeV$^3$)& $\langle m_{B\ell}^4\rangle$ 
(GeV$^4$)\\
\hline
171 & 84.42 & $7.49\times 10^3$ & $6.93\times 10^5$ & $6.65\times 10^8$ \\
\hline
173 & 84.79 & $7.55\times 10^3$ & $7.02\times 10^5$ & $6.77\times 10^8$ \\ 
\hline
175 & 85.13 & $7.61\times 10^3$ & $7.10\times 10^5$ & $6.87\times 10^8$ \\
\hline
177 & 85.45 & $7.67\times 10^3$ & $7.18\times 10^5$ & $6.97\times 10^8$ \\ 
\hline
179 & 85.77 & $7.73\times 10^3$ & $7.26\times 10^5$ & $7.06\times 10^8$ \\
\hline\end{tabular}
\end{center}
\caption{\label{pyt}
As in Table~\ref{hwt}, but using the PYTHIA event generator.} \end{table}\par

\begin{figure}[t]
\centerline{\resizebox{0.60\textwidth}{!}{\includegraphics{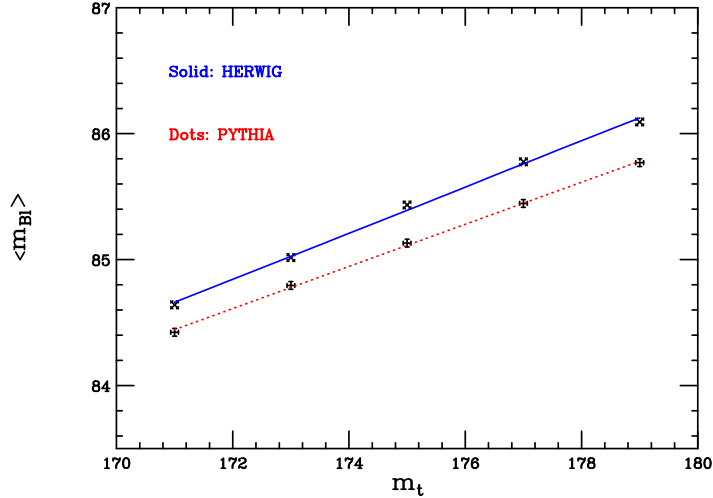}}}
\caption{Linear fits of $\langle m_{B\ell}\rangle$, as a function of 
$m_t$, using HERWIG and PYTHIA in the range 
50 GeV~$<m_{B\ell}<$~120 GeV.}
\label{trfit}
\end{figure}

As done when considering the full $m_{B\ell}$ range, we try to 
express the average value $\langle m_{B\ell}\rangle$ in terms of
$m_t$, according to a straight line.
The best linear fits read:
\begin{eqnarray}
\langle m_{B\ell}\rangle_{\mathrm{H}} &\simeq & 
53.33~\mathrm{GeV} +0.18\  m_t\ \  ;\  \ \delta = 0.034~\mathrm{GeV},\\
\langle m_{B\ell}\rangle_{\mathrm{P}} &\simeq & 
55.83~\mathrm{GeV} +0.17\  m_t\ \  ;\  \ \delta = 0.020~\mathrm{GeV}.
\end{eqnarray}
The corresponding straight lines are plotted in Fig.~\ref{trfit}.
Since the found discrepancy between HERWIG and PYTHIA
is about $\Delta \langle m_{B\ell}\rangle\simeq 200-300$~MeV, and
given the slopes of the straight lines in Fig.~\ref{trfit}, the induced 
uncertainty on the top mass, thinking of extracting it by fitting
the mean value $\langle m_{B\ell}\rangle$, goes down to 
$\Delta m_t\simeq 1.5$~GeV. It is nonetheless still a quite large 
value, well above
the estimate of the bottom-fragmentation contribution
to the Monte Carlo error given in \cite{avto}.

\section{Conclusions}
We performed a phenomenological study of bottom quark fragmentation
in top-quark decay,
using HERWIG and PYTHIA, the two most popular general-purpose Monte Carlo 
event generators.
We observed that the default parametrizations are unable
to reproduce $B$-hadron production data from ALEPH, OPAL and SLD, and 
therefore we used 
an `unofficial' fit of the hadronization cluster and string models,
following \cite{drol},
in such a way to improve the description of such data.

We used this tuning to make predictions for a few observables in 
$t\bar t$ events, depending on modelling $b$-fragmentation in top decay,
and found that the fits to $e^+e^-$ data have a remarkable impact even
on top-decay observables. Moreover, 
HERWIG and PYTHIA results still exhibit 
visible discrepancies, which depend on the different quality of the
fits to LEP and SLD data.
We studied the $B$-energy fraction in top decay, which turned out
to be roughly independent of the top mass, the $B$-lepton invariant
mass $m_{B\ell}$, exhibiting relevant dependence on $m_t$, and the
$B$ transverse momentum in the laboratory frame, which
can still be useful to determine the experimental acceptance for
$t\bar t$ events, although it
is not Lorentz-invariant and depends on the top-production mechanism
as well.

Among these quantities, we have taken particular care about $m_{B\ell}$,
whose spectrum is also interesting for the purpose of the analyses
\cite{avto,lucio}, where the top mass is reconstructed by using final states
with leptons and $J/\psi$ or muons.. 
We calculated the Mellin moments of the $m_{B\ell}$ distribution and
parametrized 
the average value $\langle m_{B\ell}\rangle$ as a linear fit of the top mass.
The found discrepancies between HERWIG and PYTHIA 
result in an uncertainty on the top mass, assuming that one can
extract it from a fit of $\langle m_{B\ell}\rangle$, which can be up to
$\Delta m_t\simeq 2$~GeV.
The Monte Carlo error due to modelling $b$-fragmentation decreases down
to $\Delta m_t\simeq 1.5$~GeV, if we restrict our analysis to the
region around the invariant-mass peak, namely 50 GeV~$<m_{B\ell}<120$~GeV.
As such estimates are quite large, our study confirms that bottom
fragmentation will
play a crucial role in top-quark analyses
at Tevatron and LHC and that having event generators reliably describing the
$b\to B$ transition will be fundamental.

A possible extension of our work clearly consists 
in employing the object-oriented versions of HERWIG and PYTHIA, 
written in C++.
In fact, the discrepancies here emphasized mainly depend on the fact that,
even after the fits, FORTRAN 
HERWIG is only marginally consistent with $B$-hadron
data from LEP and SLD. Therefore, since the preliminary results presented
in \cite{gieseke}, obtained by comparing an early version of HERWIG++
with the SLD $B$-data,
look encouraging, 
we believe that a lower $\Delta m_t$ can eventually be obtained when 
using the C++ programs. However, the analysis carried out throughout
this paper will still be valid
as long as one uses MC@NLO or matrix-element generators interfaced to
FORTRAN HERWIG and PYTHIA for showers and hadronization.

We also stressed the fact that the fits carried out in Ref.~\cite{drol} and 
reviewed in Section 2 just account for the $B$-hadron data and may spoil
the comparison with other observables, e.g., light-hadron data.
In perspective, the advanced fitting 
code Professor \cite{prof} should be a very useful
tool, as it is capable of improving the description of the $x_B$ spectrum, 
but without spoiling too much possible agreement with other data.
The use of the Professor program is currently in progress.

Furthermore, let us point out that,
whereas in our study we aimed at predicting a few top-decay
observables taking non-perturbative information from $e^+e^-$ annihilation,
possible hadron-collider data, for example on the
$m_{B\ell}$ distribution, should be very  useful to validate tools
such as HERWIG or PYTHIA. This way, one could directly fit such spectra
and tune the parameters of cluster and string models, 
without necessarily relying on the fits to the $e^+e^-$ data.
For the time being, we still 
believe that it can be nevertheless very interesting
reconsidering the studies \cite{avto,lucio}, as they strongly rely on
the Monte Carlo treatment of bottom fragmentation in top decay, and understand
whether the results on $\Delta m_t$ quoted in \cite{avto,lucio} should change
if one used the tuned versions of cluster and string models, as we did
throughout this paper.

\section*{Acknowledgements}
We are grateful to K.~Melnikov who pointed out an error in the
former version of this work.
We acknowledge discussions with L.~Cerrito, 
S.~Leone, M.L.~Mangano, M.H.~Seymour,
R. Tenchini and T.~Sj\"ostrand on these and related topics.
The work of F.M. has been supported in part by CUR Generalitat
de Catalunya under project 2009SGR502 and by the Consolider-Ingenio
2010 Program CPAN (CSD2007-00042).

\end{document}